\documentclass[aps,tightlines,superscriptaddress,groupedaddress]{revtex4}
\usepackage{graphicx}
\usepackage{dcolumn}
\usepackage{bm}
\usepackage{amssymb}
\usepackage{amsfonts}
\usepackage{amsmath}
\usepackage{epstopdf}

\begin{document}

\title{On monogamy of four-qubit entanglement}
\author{S. Shelly Sharma}
\email{shelly@uel.br}
\affiliation{Departamento de F\'{\i}sica, Universidade Estadual de Londrina, Londrina
86051-990, PR Brazil }
\author{N. K. Sharma}
\email{nsharma@uel.br}
\affiliation{Departamento de Matematica, Universidade Estadual de Londrina, Londrina
86051-990, PR Brazil }

\begin{abstract}
Our main result is a monogamy inequality satisfied by the entanglement of a
focus qubit (one-tangle) in a four-qubit pure state and entanglement of
subsystems. Analytical relations between three-tangles of three-qubit marginal states, 
two-tangles of two-qubit marginal states and unitary invariants of
four-qubit pure state are used to obtain the inequality. The contribution
of three-tangle to one-tangle is found to be half of that suggested by a
simple extension of entanglement monogamy relation for three qubits. On the
other hand, an additional contribution due to a two-qubit invariant which is
a function of three-way correlations is found. We also show that four-qubit
monogamy inequality conjecture of ref. [PRL 113, 110501 (2014)], in which
three-tangles are raised to the power $\frac{3}{2}$, does not estimate the
residual correlations, correctly, for certain subsets of four-qubit states.
A lower bound on residual four-qubit correlations is obtained.
\end{abstract} 

\maketitle

\section{Introduction}

\label{intro}

Entanglement is a necessary ingredient of any quantum computation and a
physical resource for quantum cryptography and quantum communication \cite%
{niel11}. It has also found applications in other areas such as quantum
field theory \cite{cala12}, statistical physics \cite{sahl15}, and quantum
biology \cite{lamb13}. Multipartite entanglement that comes into play in
quantum systems with more than two subsystems, is a resource for multiuser
quantum information tasks. Since the mathematical structure of multipartite
states is much more complex than that of bipartite states, the
characterization of multipartite entanglement is a far more challenging task 
\cite{horo09}.

Monogamy is a unique feature of quantum entanglement, which determines how
entanglement is distributed amongst the subsystems. Three-qubit entanglement
is known to satisfy a quantitative constraint, known as CKW monogamy
inequality \cite{coff00}. In recent articles \cite{regu14,regu15,regu16}, it
has been shown that the most natural extension of CKW inequality to
four-qubit entanglement is violated by some of the four-qubit states and
different ways to extend the monogamy inequality to four-qubits have been
conjectured. For a subclass of four-qubit generic states, an extension of
strong monogamy inequality to negativity and squared negativity \cite{karm16}
is satisfied, however, there exist four-qubit states for which negativity
and squared negativity are not strongly monogamous. Three-qubit states show
two distinct types of entanglement. As we go to four qubits, additional
degrees of freedom make it possible for new entanglement types to emerge. It
is signalled by the fact that corresponding to the three-qubit invariant
that detects genuine three-way entanglement of a three-qubit pure state, a
four-qubit pure state has five three-qubit invariants for each set of three
qubits \cite{shar16}. An $m$-qubit invariant is understood to be a function
of state coefficients which remains invariant under the action of a local
unitary transformation on the state of any one of the $m$ qubits. A valid
discussion of entanglement monogamy for four qubits, therefore, must include
contributions from invariants that detect new entanglement types.

This article is an attempt to identify, analytically, the contributions of
two-tangles (pairwise entanglement), three-tangles (genuine three-way
entanglement) and four tangles to entanglement of a focus qubit with the
three remaining qubits (one-tangle) in a four qubit state. To do this, we
express one-tangle in terms of two-qubit invariants. Monogamy inequality
constraint on four qubit entanglement is obtained by comparing the
one-tangle with upper bounds on two-tangles and three-tangles \cite{shar216}
defined on two and three qubit marginal states. Contribution of
three-tangles to one-tangle is found to be half of what is expected from a
direct generalization of CKW inequality to four qubits. The difference
arises due to new entanglement modes that are available to four qubits. It
is verified that the "residual entanglement", obtained after subtracting the
contributions of two-tangles and three-tangles from one-tangle, is greater
or equal to genuine four-tangle. Genuine four-tangle \cite{shar14,shar16} is
a degree eight function of state coefficients of the pure state. Besides
that, the "residual entanglement" also contains contributions from square of
degree-two four-tangle \cite{shar10,luqu03} and degree-four invariants which
quantify the entanglement of a given pair of qubits with its complement in a
four-qubit pure state \cite{shar10}.

\section{ One-tangle of a focus qubit in a four-qubit State}

\label{tangle}

We start by expressing one-tangle of a focus qubit in a four-qubit state in
terms of two-qubit invariant functions of state coefficients. Entanglement
of qubit $A_{1}$ with $A_{2}$ in a two-qubit pure state 
\begin{equation}
\left\vert \Psi _{12}\right\rangle
=\sum_{i_{1},i_{2}}a_{i_{1}i_{2}}\left\vert i_{1}i_{2}\right\rangle ;\quad
(i_{m}=0,1)
\end{equation}%
is quantified by two-tangle defined as 
\begin{equation}
\tau _{1|2}\left( \left\vert \Psi _{12}\right\rangle \right) =2\left\vert
D^{00}\right\vert ,
\end{equation}%
where $D^{00}=a_{00}a_{11}-a_{10}a_{01}$ is a two-qubit invariant. Here $%
a_{i_{1}i_{2}}$ are the state coefficients. On a four qubit pure state,
however, for each choice of a pair of qubits one identifies nine two-qubit
invariants. Three-tangle \cite{coff00} of a three-qubit pure state is
defined in terms of modulus of a three-qubit invariant. On the most general
four qubit state, on the other hand, we have five three-qubit invariants
corresponding to a given set of three qubits. Four-qubit invariant that
quantifies the sum of three-way and four-way correlations of a three-qubit
partition in a pure state is known to be a degree-eight invariant \cite%
{shar16}, which is a function of three-qubit invariants. It is natural to
expect that the monogamy inequality for four qubits takes into account the
entanglement modes available exclusively to four-qubit system. To
understand, how various two-tangles and three-tangles add up to generate
total entanglement of a focus qubit in a pure four-qubit state, we follow
the steps listed below:

(1) Write down one-tangle of focus qubit as a sum of two-qubit invariants.

(2) Express two-tangles, three-tangles and four-tangle or the upper bounds
on the tangles in terms of two-qubit invariants.

(3) Rewrite one-tangle in terms of tangles defined on two- and three-qubit
reduced states and \ "residual four-qubit correlations".

(4) Compare the "residual four-qubit correlations" with the lower bound on
four-qubit correlations written in terms of four-qubit invariants.

To facilitate the identification of two-qubit and three-qubit invariants, we
use the formalism of determinants of two by two matrices of state
coefficients referred to as negativity fonts. For more on definition and
physical meaning of determinants of negativity fonts, please refer to
section (VI) of ref. \cite{shar16}.

For the purpose of this article, we write down and use the determinants of
negativity fonts of a four-qubit state when qubit $A_{1}$ is the focus
qubit. On a four-qubit pure state, written as%
\begin{equation}
\left\vert \Psi _{1234}\right\rangle
=\sum_{i_{1},i_{2},i_{3}i_{4}}a_{i_{1}i_{2}i_{3}i_{4}}\left\vert
i_{1}i_{2}i_{3}i_{4}\right\rangle ,\quad \left( i_{m}=0,1\right) ,
\label{4state}
\end{equation}%
where state coefficients $a_{i_{1}i_{2}i_{3}i_{4}}$ are complex numbers and $%
i_{m}\ $refers to the basis state of qubit $A_{m}$, $\left( m=1,2,3\right) $%
, we identify the determinants of two-way negativity fonts to be $D_{\left(
A_{3}\right) _{i_{3}}\left( A_{4}\right)
_{i_{4}}}^{00}=a_{00i_{3}i_{4}}a_{11i_{3}i_{4}}-a_{10i_{3}i_{4}}a_{01i_{3}i_{4}} 
$, $D_{\left( A_{2}\right) _{i_{2}}\left( A_{4}\right)
_{i_{4}}}^{00}=a_{0i_{2}0i_{4}}a_{1i_{2}1i_{4}}-a_{1i_{2}0i_{4}}a_{0i_{2}1i_{4}} 
$, and $D_{\left( A_{2}\right) _{i_{2}}\left( A_{3}\right)
_{i_{3}}}^{00}=a_{0i_{2}i_{3}0}a_{1i_{2}i_{3}1}-a_{1i_{2}i_{3}0}a_{0i_{2}i_{3}1} 
$. Besides that we also have $D_{\left( A_{4}\right)
_{i_{4}}}^{00i_{3}}=a_{00i_{3}i_{4}}a_{11,i_{3}\oplus
1,i_{4}}-a_{10i_{3}i_{4}}a_{01,i_{3}\oplus 1,i_{4}}$ (three-way), $D_{\left(
A_{3}\right) _{i_{3}}}^{00i_{4}}=a_{00i_{3}i_{4}}a_{11i_{3},i_{4}\oplus
1}-a_{10i_{3}i_{4}}a_{01i_{3},i_{4}\oplus 1}$ (three-way), $D_{\left(
A_{2}\right) _{i_{2}}}^{00i_{4}}=a_{0i_{2}0i_{4}}a_{1i_{2}1i_{4}\oplus
1}-a_{1i_{2}0i_{4}}a_{0i_{2}1i_{4}\oplus 1}$ (three-way), and $%
D^{00i_{3}i_{4}}=a_{00i_{3}i_{4}}a_{11,i_{3}\oplus 1,i_{4}\oplus
1}-a_{10i_{3}i_{4}}a_{01,i_{3}\oplus 1,i_{4}\oplus 1}$ (four-way), as the
determinants of negativity fonts.

One-tangle given by $\tau _{1|234}\left( \left\vert \Psi
_{1234}\right\rangle \right) =4\det \left( \rho _{1}\right) $, where $\rho
_{1}=$Tr$_{A_{2}A_{3}A_{4}}\left( \left\vert \Psi _{1234}\right\rangle
\left\langle \Psi _{1234}\right\vert \right) $, quantifies the entanglement
of qubit $A_{1}$ with $A_{2}$, $A_{3}$ and $A_{4}$. It is four times the
square of negativity of partial transpose of four-qubit pure state with
respect to qubit $A_{1}$ \cite{pere96}. Negativity, in general, does not
satisfy the monogamy relation. However, it has been shown by He and Vidal 
\cite{he15} that negativity can satisfy monogamy relation in the setting
provided by disentangling theorem. It is easily verified that%
\begin{eqnarray}
\tau _{1|234}\left( \left\vert \Psi _{1234}\right\rangle \right) &=&4\left[
\sum\limits_{i_{3},i_{4}=0}^{1}\left\vert D_{\left( A_{3}\right)
_{i_{3}}\left( A_{4}\right) _{i_{4}}}^{00}\right\vert
^{2}+\sum\limits_{i_{2},i_{4}=0}^{1}\left\vert D_{\left( A_{2}\right)
_{i_{2}}\left( A_{4}\right) _{i_{4}}}^{00}\right\vert
^{2}+\sum\limits_{i_{2},i_{3}=0}^{1}\left\vert D_{\left( A_{2}\right)
_{i_{2}}\left( A_{3}\right) _{i_{3}}}^{00}\right\vert ^{2}\right.  \notag \\
&&\left. +\sum\limits_{i_{3},i_{4}=0}^{1}\left\vert D_{\left( A_{4}\right)
_{i_{4}}}^{00i_{3}}\right\vert
^{2}+\sum\limits_{i_{3},i_{4}=0}^{1}\left\vert D_{\left( A_{3}\right)
_{i_{3}}}^{00i_{4}}\right\vert
^{2}+\sum\limits_{i_{2},i_{4}=0}^{1}\left\vert D_{\left( A_{2}\right)
_{i_{2}}}^{00i_{4}}\right\vert
^{2}+\sum\limits_{i_{3},i_{4}=0}^{1}\left\vert D^{00i_{3}i_{4}}\right\vert
^{2}\right] .  \label{tanglea1}
\end{eqnarray}%
One-tangle depends on $2-$way, $3-$way and $4-$way correlations of focus
qubit $A_{1}$ with the rest of the system.

\section{Definitions of two-tangles and three-tangle}

\label{threeq}

This section contains the definitions of two-tangles and three-tangles for
pure and mixed three-qubit states. Consider a three-qubit pure state%
\begin{equation}
\left\vert \Psi _{123}\right\rangle
=\sum\limits_{i_{1},i_{2},i_{3}}a_{i_{1}i_{2}i_{3}}\left\vert
i_{1}i_{2}i_{3}\right\rangle ,\quad i_{m}=0,1,  \label{3qstate}
\end{equation}%
Using the notation from ref. \cite{shar16}, we define $D_{\left(
A_{3}\right) _{i_{3}}}^{00}=a_{00i_{3}}a_{11i_{3}}-a_{10i_{3}}a_{01i_{3}},$ (%
$i_{3}=0,1$) (the determinant of a two-way negativity font) and $%
D^{00i_{3}}=a_{00i_{3}}a_{11i_{3}+1}-a_{10i_{3}}a_{01i_{3}+1}$, ($i_{3}=0,1$%
) (the determinant of a three-way negativity font). Entanglement of qubit $%
A_{1}$ with the rest of the system is quantified by one-tangle $\tau
_{1|23}\left( \left\vert \Psi _{123}\right\rangle \right) =4\det \left( \rho
_{1}\right) $, where $\rho _{1}=$Tr$_{A_{2}A_{3}}\left( \left\vert \Psi
_{123}\right\rangle \left\langle \Psi _{123}\right\vert \right) $. One can
verify that%
\begin{equation}
\tau _{1|23}\left( \left\vert \Psi _{123}\right\rangle \right)
=4\sum\limits_{i_{3}=0}^{1}\left\vert D_{\left( A_{3}\right)
_{i_{3}}}^{00}\right\vert ^{2}+4\sum\limits_{i_{3}=0}^{1}\left\vert
D^{00i_{3}}\right\vert ^{2}+4\sum\limits_{i_{2}=0}^{1}\left\vert D_{\left(
A_{2}\right) _{i_{2}}}^{00}\right\vert ^{2}.  \label{tau123}
\end{equation}%
For qubit pair $A_{1}A_{2}$ in $\left\vert \Psi _{123}\right\rangle $, we
identify three two-qubit invariants that is 
\begin{equation}
D_{\left( A_{3}\right) _{0}}^{00},\frac{\left( D^{000}+D^{001}\right) }{2}%
,D_{\left( A_{3}\right) _{1}}^{00}  \label{a1a2qinv}
\end{equation}%
while for the pair $A_{1}A_{3}$ two-qubit invariants are 
\begin{equation}
D_{\left( A_{2}\right) _{0}}^{00},\frac{\left( D^{000}-D^{001}\right) }{2}%
,D_{\left( A_{2}\right) _{1}}^{00}.  \label{a1a3qinv}
\end{equation}%
These two-qubit invariants transform under a unitary $U=\frac{1}{\sqrt{%
1+\left\vert x\right\vert ^{2}}}\left[ 
\begin{array}{cc}
1 & -x^{\ast } \\ 
x & 1%
\end{array}%
\right] $ on the third qubit in a way analogous to the complex functions $%
A(x)$, $B(x)$ and $C(x)$ of Appendix \ref{A1}. Then the invariants
corresponding to $I^{(1)}$, $I^{(2)}$ and $\tau ^{2}$ are three-qubit
invariants for the given choice of qubit pair. In Table I, we enlist the
correspondence of two-qubit invariants for qubit pairs $A_{1}A_{2}$ and $%
A_{1}A_{3}$, with complex numbers $A$, $B$ and $C$ of Appendix \ref{A1} and
set the notation for invariants corresponding to $I^{(1)}$, $I^{(2)}$ and $%
\tau ^{2}.$

\begin{table}[th]
\caption{Two-tangles and three-tangle in terms of two-qubit invariants of a
three-qubit pure state. Here $i=0,1$, $I^{(1)}=\left\vert A\right\vert ^{2}+%
\frac{1}{2}\left\vert B\right\vert ^{2}+\left\vert C\right\vert ^{2}$ , $%
I^{(2)}=\left\vert B^{2}-4AC\right\vert $ and $I=4I^{(1)}-2I^{(2)}$
(Appendix \protect\ref{A1}). }\centering
\par
\begin{tabular}{||c||c||c||c||c||c||c||c||}
\hline\hline
& \multicolumn{3}{||c}{Two-qubit invariants} & \multicolumn{4}{||c||}{
Three-qubit invariants} \\ \hline\hline
$%
\begin{array}{c}
\text{Qubit} \\ 
\text{Pair}%
\end{array}%
$ & $A$ & $B$ & $C$ & $I^{(1)}$ & $I^{(2)}$ & $4I^{(2)}$ & $\tau ^{2}$ \\ 
\hline\hline
$A_{1}A_{2}$ & $D_{\left( A_{3}\right) _{0}}^{00}$ & $D^{000}+D^{001}$ & $%
D_{\left( A_{3}\right) _{1}}^{00}$ & $N_{A_{3}}$ & $\left\vert I_{3,4}\left(
\left\vert \Psi _{123}\right\rangle \right) \right\vert $ & $\tau
_{1|2|3}\left( \left\vert \Psi _{123}\right\rangle \right) $ & $\left[ \tau
_{1|2}\left( \rho _{12}\right) \right] ^{2}$ \\ \hline\hline
$A_{1}A_{3}$ & $D_{\left( A_{2}\right) _{0}}^{00}$ & $D^{000}-D^{001}$ & $%
D_{\left( A_{2}\right) _{1}}^{00}$ & $N_{A_{2}}$ & $\left\vert I_{3,4}\left(
\left\vert \Psi _{123}\right\rangle \right) \right\vert $ & $\tau
_{1|2|3}\left( \left\vert \Psi _{123}\right\rangle \right) $ & $\left[ \tau
_{1|3}\left( \rho _{13}\right) \right] ^{2}$ \\ \hline\hline
\end{tabular}%
\end{table}

One-tangle in terms of three-qubit invariants listed in column five of Table
I reads as%
\begin{equation}
\tau _{1|23}\left( \left\vert \Psi _{123}\right\rangle \right)
=4N_{A_{3}}+4N_{A_{2}}.
\end{equation}%
Three tangle \cite{coff00} of pure state $\left\vert \Psi
_{123}\right\rangle $ is equal to the modulus of the polynomial invariant of
degree four that is%
\begin{equation*}
\tau _{1|2|3}\left( \left\vert \Psi _{123}\right\rangle \right) =4\left\vert
I_{3,4}\left( \left\vert \Psi _{123}\right\rangle \right) \right\vert ,
\end{equation*}%
where%
\begin{eqnarray}
I_{3,4}\left( \left\vert \Psi _{123}\right\rangle \right) &=&\left(
D^{000}+D^{001}\right) ^{2}-4D_{\left( A_{3}\right) _{0}}^{00}D_{\left(
A_{3}\right) _{1}}^{00}  \notag \\
&=&\left( D^{000}-D^{001}\right) ^{2}-4D_{\left( A_{2}\right)
_{0}}^{00}D_{\left( A_{2}\right) _{1}}^{00}.  \label{three_invariant}
\end{eqnarray}%
The entanglement measure $\tau _{1|2|3}\left( \left\vert \Psi
_{123}\right\rangle \right) $ is extended to a mixed state of three qubits
via convex roof extension that is%
\begin{equation}
\left[ \tau _{1|2|3}\left( \rho _{123}\right) \right] ^{\frac{1}{2}%
}=\min_{\left\{ p_{i},\left\vert \phi _{123}^{\left( i\right) }\right\rangle
\right\} }\sum\limits_{i}p_{i}\left[ \tau _{1|2|3}\left( \left\vert \phi
_{123}^{\left( i\right) }\right\rangle \right) \right] ^{\frac{1}{2}},
\label{mix3tangle}
\end{equation}%
where minimization is taken over all complex decompositions $\left\{
p_{i},\left\vert \phi _{123}^{\left( i\right) }\right\rangle \right\} $ of $%
\rho _{123}$. Here $p_{i}$ is the probability of finding the normalized
state $\left\vert \phi _{123}^{\left( i\right) }\right\rangle $ in the mixed
state $\rho _{123}.$

Two-tangle of the state $\rho _{12}=\sum\limits_{i}p_{i}\left\vert \phi
_{12}^{\left( i\right) }\right\rangle \left\langle \phi _{12}^{\left(
i\right) }\right\vert $ is constructed through convex roof extension as%
\begin{equation}
\tau _{1|2}\left( \rho _{12}\right) =2\min_{\left\{ p_{i},\phi _{12}^{\left(
i\right) }\right\} }\sum\limits_{i}p_{i}\left\vert D^{00}\left( \left\vert
\phi _{12}^{\left( i\right) }\right\rangle \right) \right\vert .
\label{mix2tangle}
\end{equation}%
Two-tangle $\tau _{1|2}\left( \rho _{12}\right) =C\left( \rho _{12}\right) $%
, where $C\left( \rho _{12}\right) $ is the concurrence \cite{hill97,woot98}%
. One can verify that the invariants $N_{A_{3}}$, $N_{A_{2}}$, corresponding
two-tangles, and three-tangle saturate the inequalities corresponding to Eq.
(\ref{iineq}) that is%
\begin{equation}
4N_{A_{3}}=\left[ \tau _{1|2}\left( \rho _{12}\right) \right] ^{2}+\frac{1}{2%
}\tau _{1|2|3}\left( \left\vert \Psi _{123}\right\rangle \right) ,
\label{na3}
\end{equation}%
and 
\begin{equation}
4N_{A_{2}}=\left[ \tau _{1|3}\left( \rho _{13}\right) \right] ^{2}+\frac{1}{2%
}\tau _{1|2|3}\left( \left\vert \Psi _{123}\right\rangle \right) .
\label{na2}
\end{equation}%
Since $\tau _{1|23}\left( \left\vert \Psi _{123}\right\rangle \right)
=4\left( N_{A_{2}}+N_{A_{3}}\right) $, we obtain%
\begin{equation}
\tau _{1|23}\left( \left\vert \Psi _{123}\right\rangle \right) \geq \left[
\tau _{1|2}\left( \rho _{12}\right) \right] ^{2}+\left[ \tau _{1|3}\left(
\rho _{13}\right) \right] ^{2},  \label{ckw}
\end{equation}%
which is the well known CKW inequality. From Eqs. (\ref{na3}) and (\ref{na2}%
), the distribution of entanglement in a three-qubit state and its two-qubit
marginals satisfies the following relation:%
\begin{equation}
\tau _{1|23}\left( \left\vert \Psi _{123}\right\rangle \right) =\left[ \tau
_{1|2}\left( \rho _{12}\right) \right] ^{2}+\left[ \tau _{1|3}\left( \rho
_{13}\right) \right] ^{2}+\tau _{1|2|3}\left( \left\vert \Psi
_{123}\right\rangle \right) .  \label{onetangle3q}
\end{equation}

Moduli of two-qubit invariants, which depend only on the determinants of
three-way negativity fonts, are used to define new two-tangles on the state $%
\rho _{123}=\sum\limits_{i}p_{i}\left\vert \phi _{123}^{\left( i\right)
}\right\rangle \left\langle \phi _{123}^{\left( i\right) }\right\vert $ via%
\begin{equation}
\tau _{1|p}^{\left( new\right) }\left( \rho _{123}\right) =\min_{\left\{
p_{i},\phi _{123}^{\left( i\right) }\right\} }\sum\limits_{i}p_{i}\left(
2\left\vert T_{1p}\left( \left\vert \phi _{123}^{\left( i\right)
}\right\rangle \right) \right\vert \right) ,  \label{1pnew}
\end{equation}%
where%
\begin{equation}
T_{12}\left( \left\vert \Psi _{123}\right\rangle \right) =D^{000}\left(
\left\vert \Psi _{123}\right\rangle \right) +D^{001}\left( \left\vert \Psi
_{123}\right\rangle \right) ,  \label{t12}
\end{equation}%
and%
\begin{equation}
T_{13}\left( \left\vert \Psi _{123}\right\rangle \right) =D^{000}\left(
\left\vert \Psi _{123}\right\rangle \right) -D^{001}\left( \left\vert \Psi
_{123}\right\rangle \right) .  \label{t13}
\end{equation}

\subsection{What does $\protect\tau _{1|p}^{\left( new\right) }\left( 
\protect\rho _{123}\right) $ measure?}

To understand the correlations represented by $\tau _{1|p}^{\left(
new\right) }\left( \rho _{123}\right) $, we examine a generic three-qubit
state in its canonical form. A state is said to be in the canonical form
when it is expressed as a superposition of minimal number of local basis
product states (LBPS) \cite{acin01}. The state coefficients of this form
carry all the information about the non-local properties of the state, and
do so minimally. Starting from a generic state in the basis $\left\vert
i_{1}i_{2}i_{3}\right\rangle $ (Eq. (\ref{3qstate})), local unitary
transformations allow us to write it in a form with the minimal number of
LBPS.\ As a first step towards writing the state $\left\vert \Psi
_{123}\right\rangle $ in canonical form with respect to $r^{th}$ qubit, we
chose a unitary that results in a state on which one of the two-way
two-qubit invariants is zero that is $D_{\left( A_{r}\right) _{0}}^{00}=0$
or $D_{\left( A_{r}\right) _{1}}^{00}=0$. For example a unitary $U^{3}=\frac{%
1}{\sqrt{1+\left\vert x\right\vert ^{2}}}\left[ 
\begin{array}{cc}
1 & -x^{\ast } \\ 
x & 1%
\end{array}%
\right] ;$ with $x^{\ast }=\frac{\left( D^{000}+D^{001}\right) \pm \sqrt{%
I_{3,4}}}{2D_{\left( A_{3}\right) _{1}}^{00}}$, acting on qubit $A_{3}$
gives a state U$^{3}\left\vert \Psi _{123}\right\rangle
=\sum\limits_{i_{1},i_{2},i_{3}}b_{i_{1}i_{2}i_{3}}\left\vert
i_{1}i_{2}i_{3}\right\rangle $ , such that $b_{000}b_{110}-b_{100}b_{010}=0$%
. After eliminating $b_{100}$, the state re \ ads as%
\begin{equation}
U^{3}\left\vert \Psi _{123}\right\rangle =b_{000}\left( \left\vert
0\right\rangle _{1}+\frac{b_{110}}{b_{010}}\left\vert 1\right\rangle
_{1}\right) \left( \left\vert 0\right\rangle _{2}+\frac{b_{010}}{b_{000}}%
\left( \left\vert 1\right\rangle _{2}\right) \right) \left\vert
0\right\rangle _{3}+\sum\limits_{i_{1},i_{2}}b_{i_{1}i_{2}i_{3}}\left\vert
i_{1}i_{2}1\right\rangle .
\end{equation}%
It is straight forward to write down the local unitaries $U^{1}$ and $U^{2}$
that lead to the canonical form, 
\begin{equation}
\left\vert \Psi _{123}\right\rangle _{c}=c_{000}\left\vert 000\right\rangle
+c_{001}\left\vert 001\right\rangle +c_{101}\left\vert 101\right\rangle
+c_{011}\left\vert 011\right\rangle +c_{111}\left\vert 111\right\rangle 
\text{.}  \label{psicanon}
\end{equation}%
Notice that on canonical state, $\left[ 4\left\vert T_{12}\left( \left\vert
\Psi _{123}\right\rangle _{c}\right) \right\vert \right] ^{2}=\left[
4\left\vert T_{13}\left( \left\vert \Psi _{123}\right\rangle _{c}\right)
\right\vert \right] ^{2}=\tau _{1|2|3}\left( \left\vert \Psi
_{123}\right\rangle \right) $. Next we determine the range of values that $%
\left[ 4\left\vert T_{12}\left( \left\vert \Psi _{123}\right\rangle \right)
\right\vert \right] ^{2}$ takes on a generic state $\left\vert \Psi
_{123}\right\rangle $.

Combining the definition of $N_{A_{3}}$ (from Table I) for the state $%
\left\vert \Psi _{123}\right\rangle $, with the result of Eq. (\ref{na3}),
that is%
\begin{eqnarray}
4N_{A_{3}} &=&4\left\vert D_{\left( A_{3}\right) _{0}}^{00}\right\vert
^{2}+2\left\vert D^{000}+D^{001}\right\vert ^{2}+4\left\vert D_{\left(
A_{3}\right) _{1}}^{00}\right\vert ^{2}  \notag \\
&=&\left[ \tau _{1|2}\left( \rho _{12}\right) \right] ^{2}+\frac{1}{2}\tau
_{1|2|3}\left( \left\vert \Psi _{123}\right\rangle \right) ,
\end{eqnarray}%
we obtain%
\begin{equation}
4\left\vert T_{12}\left( \left\vert \Psi _{123}\right\rangle \right)
\right\vert ^{2}=\tau _{1|2|3}\left( \left\vert \Psi _{123}\right\rangle
\right) -D_{c}\ ,  \label{t12dc}
\end{equation}%
where%
\begin{equation*}
D_{c}=8\left\vert D_{\left( A_{3}\right) _{0}}^{00}\right\vert
^{2}+8\left\vert D_{\left( A_{3}\right) _{1}}^{00}\right\vert ^{2}-2\left[
\tau _{1|2}\left( \rho _{12}\right) \right] ^{2}.
\end{equation*}%
Since $4\left( \left\vert D_{\left( A_{3}\right) _{0}}^{00}\right\vert
^{2}+\left\vert D_{\left( A_{3}\right) _{1}}^{00}\right\vert ^{2}\right)
\geq \left[ \tau _{1|2}\left( \rho _{12}\right) \right] ^{2}$ ($D_{c}\geq 0$%
), the value of $\left\vert T_{12}\left( \left\vert \Psi _{123}\right\rangle
\right) \right\vert $ satisfies%
\begin{equation}
\tau _{1|2|3}\left( \left\vert \Psi _{123}\right\rangle \right) \geq
4\left\vert T_{12}\left( \left\vert \Psi _{123}\right\rangle \right)
\right\vert ^{2}\geq 0.  \label{t12newcondition}
\end{equation}%
In general, the difference $D_{c}=\tau _{p|q|r}\left( \left\vert \Psi
_{pqr}\right\rangle \right) -4\left\vert T_{pq}\left( \left\vert \Psi
_{pqr}\right\rangle \right) \right\vert ^{2}$ measures the distance of a
given three-qubit state from its canonical form with respect to $r^{th}$
qubit.

On a pure state of three qubits, $\tau _{1|2}^{\left( new\right) }\left(
\left\vert \Psi _{123}\right\rangle \right) =\min $ $\left\vert T_{12}\left(
U^{3}\left\vert \Psi _{123}\right\rangle \right) \right\vert =0$. The state
on which $\left\vert T_{12}\left( U^{3}\left\vert \Psi _{123}\right\rangle
\right) \right\vert =0$, is obtained by a unitary transformation $U^{3}=%
\frac{1}{\sqrt{1+\left\vert x\right\vert ^{2}}}\left[ 
\begin{array}{cc}
1 & -x^{\ast } \\ 
x & 1%
\end{array}%
\right] ,$ such that%
\begin{equation}
x^{\ast }=\frac{\left\vert D_{\left( A_{3}\right) _{0}}^{00}\right\vert
^{2}-\left\vert D_{\left( A_{3}\right) _{1}}^{00}\right\vert ^{2}\pm \frac{1%
}{4}\sqrt{J_{0}}}{D_{\left( A_{3}\right) _{1}}^{00}\left(
D^{000}+D^{001}\right) ^{\ast }+\left( D_{\left( A_{3}\right)
_{0}}^{00}\right) ^{\ast }\left( D^{000}+D^{001}\right) },
\end{equation}%
where three-qubit invariant $J_{0}$ reads as%
\begin{equation}
J_{0}=\left[ \tau _{1|2}\left( \rho _{12}\right) \right] ^{2}\left[ \left[
\tau _{1|2}\left( \rho _{12}\right) \right] ^{2}+\tau _{1|2|3}\left(
\left\vert \Psi _{123}\right\rangle \right) \right] .
\end{equation}

Next, consider the three-qubit mixed state $\rho
_{123}=\sum\limits_{i=0,1}\left\vert \Phi _{123}^{\left( i\right)
}\right\rangle \left\langle \Phi _{123}^{\left( i\right) }\right\vert $,
where $\left\vert \Phi _{123}^{\left( i\right) }\right\rangle $ is an
un-normalized state. Let the set of two-qubit invariants for the pair $%
A_{1}A_{2}$ in the state $\left\vert \Phi _{123}^{\left( i\right)
}\right\rangle $ be%
\begin{equation}
\left( D_{\left( A_{3}\right) _{0}}^{00}\right) ^{\left( i\right) },\frac{1}{%
2}\left( D^{000}+D^{001}\right) ^{\left( i\right) },\left( D_{\left(
A_{3}\right) _{1}}^{00}\right) ^{\left( i\right) }.
\end{equation}%
New two-qubit invariant (Eq. (\ref{1pnew})) on $\rho _{123}$ is given by%
\begin{equation}
\tau _{1|2}^{\left( new\right) }\left( \rho _{123}\right) =2\min_{\left\{
\left\vert \Phi _{123}^{\left( i\right) }\right\rangle \right\} }\left[
\left\vert T_{12}\left( \left\vert \Phi _{123}^{\left( 0\right)
}\right\rangle \right) \right\vert +\left\vert T_{12}\left( \left\vert \Phi
_{123}^{\left( 1\right) }\right\rangle \right) \right\vert \right] \text{.}
\end{equation}%
where from Eq. (\ref{t12dc}),

\begin{equation}
4\left\vert T_{12}\left( \left\vert \Phi _{123}^{\left( i\right)
}\right\rangle \right) \right\vert ^{2}=\tau _{1|2|3}\left( \left\vert \Phi
_{123}^{\left( i\right) }\right\rangle \right) -D_{c}^{\left( i\right) },
\end{equation}%
with $D_{c}^{\left( i\right) }$ defined as%
\begin{equation}
D_{c}^{\left( i\right) }=8\left\vert \left( D_{\left( A_{3}\right)
_{0}}^{00}\right) ^{\left( i\right) }\right\vert ^{2}+8\left\vert \left(
D_{\left( A_{3}\right) _{1}}^{00}\right) ^{\left( i\right) }\right\vert
^{2}-2\left[ \tau _{1|2}\left( \left\vert \Phi _{123}^{\left( i\right)
}\right\rangle \right) \right] ^{2}.
\end{equation}%
If $U^{3}$ and $V^{3}$ are the local unitaries on the third qubit such that $%
T_{12}\left( \left\vert U^{3}\Phi _{123}^{\left( 0\right) }\right\rangle
\right) =0$ and $T_{12}\left( \left\vert V^{3}\Phi _{123}^{\left( 1\right)
}\right\rangle \right) =0$, then%
\begin{equation}
\tau _{1|2}^{\left( new\right) }\left( \rho _{123}\right) =2\min \left\{
\left\vert T_{12}\left( \left\vert U^{3}\Phi _{123}^{\left( 1\right)
}\right\rangle \right) \right\vert ,\left\vert T_{12}\left( \left\vert
V^{3}\Phi _{123}^{\left( 0\right) }\right\rangle \right) \right\vert
\right\} \text{.}  \label{mintau12}
\end{equation}%
Obviously, the value of $\tau _{1|2}^{\left( new\right) }\left( \rho
_{123}\right) $ satisfies either the condition%
\begin{equation}
\tau _{1|2|3}\left( \left\vert \Phi _{123}^{\left( 0\right) }\right\rangle
\right) \geq \left[ \tau _{1|2}^{\left( new\right) }\left( \left\vert \rho
_{123}\right\rangle \right) \right] ^{2}\geq 0,  \label{con1}
\end{equation}%
or the constraint%
\begin{equation}
\tau _{1|2|3}\left( \left\vert \Phi _{123}^{\left( 1\right) }\right\rangle
\right) \geq \left[ \tau _{1|2}^{\left( new\right) }\left( \left\vert \rho
_{123}\right\rangle \right) \right] ^{2}\geq 0.  \label{con2}
\end{equation}

\section{Tangles and Three-qubit invariants of a four-qubit state}

\label{tangles4q}

In this section, we identify relevant combinations of two-qubit invariants
that remain invariant under a local unitary on the third qubit. Three-qubit
invariants that we look for are the ones related to tangles of three-qubit
reduced states obtained from four-qubit pure state by tracing out the
degrees of freedom of the fourth qubit. For any given pair of qubits in a
general four-qubit state, there are nine two-qubit invariants. Of the six
degree-four three-qubit invariants constructed from the set of nine
two-qubit invariants, one is defined only on the pure state. Five remaining
invariants are functions of three-tangles and two-tangles. In Table II, we
identify sets of two-qubit invariants of a four-qubit state which transform
under a unitary, $U=\frac{1}{\sqrt{1+\left\vert x\right\vert ^{2}}}\left[ 
\begin{array}{cc}
1 & -x^{\ast } \\ 
x & 1%
\end{array}%
\right] $ on the third qubit in the same way as the functions $A(x)$, $B(x)$
and $C(x)$ of Appendix \ref{A1}.

\begin{table}[th]
\caption{ Two-tangles and three-tangles in terms of two-qubit invariants of
a four-qubit pure state. Here $i=0,1$, {$I^{(1)}=\left\vert A\right\vert
^{2}+\frac{1}{2}\left\vert B\right\vert ^{2}+\left\vert C\right\vert ^{2}$, $%
I^{(2)}=$}$\left\vert {B^{2}-4AC}\right\vert ${\ and }${\protect\tau ^{2}}${$%
=4I^{(1)}-2I^{(2)}$} (Appendix \protect\ref{A1}). }\centering%
\begin{tabular}{||l||l||l||l||l||l||l||}
\hline\hline
& \multicolumn{3}{||l||}{Two-qubit invariants} & \multicolumn{3}{||l||}{
Three-qubit invariants} \\ \hline\hline
$%
\begin{array}{c}
\text{Qubit} \\ 
\text{pair}%
\end{array}%
\downarrow $ & $A$ & $B$ & $C$ & $I^{(1)}$ & $I^{(2)}$ & $\tau ^{2}$ \\ 
\hline\hline
$A_{1}A_{2}$ & $D_{\left( A_{3}\right) _{0}\left( A_{4}\right) _{i}}^{00}$ & 
$D_{\left( A_{4}\right) _{i}}^{000}+D_{\left( A_{4}\right) _{i}}^{001}$ & $%
D_{\left( A_{3}\right) _{1}\left( A_{4}\right) _{i}}^{00}$ & $%
N_{A_{4}}^{\left( i\right) }$ & $\left\vert I_{3,4}\left( \left\vert \Phi
_{123}^{\left( i\right) }\right\rangle \right) \right\vert $ & $\left[ \tau
_{1|2}\left( \left\vert \Phi _{123}^{\left( i\right) }\right\rangle \right) %
\right] ^{2}$ \\ \hline\hline
$A_{1}A_{2}$ & $D_{\left( A_{3}\right) _{0}}^{000}+D_{\left( A_{3}\right)
_{0}}^{001}$ & $\left( 
\begin{array}{c}
D^{0000}+D^{0010} \\ 
+D^{0001}+D^{0011}%
\end{array}%
\right) $ & $D_{\left( A_{3}\right) _{1}}^{000}+D_{\left( A_{3}\right)
_{1}}^{001}$ & $M_{A_{3}}$ & $\left\vert \left( I_{3}\right)
_{A_{4}}^{\left( new\right) }\left( \left\vert \Psi _{1234}\right\rangle
\right) \right\vert $ & $\left[ \tau _{1|2}^{(new)}\left( \rho _{124}\right) %
\right] ^{2}$ \\ \hline\hline
$A_{1}A_{3}$ & $D_{\left( A_{2}\right) _{i}\left( A_{4}\right) _{0}}^{00}$ & 
$D_{\left( A_{2}\right) _{i}}^{000}+D_{\left( A_{2}\right) _{i}}^{001}$ & $%
D_{\left( A_{2}\right) _{i}\left( A_{4}\right) _{1}}^{00}$ & $%
N_{A_{2}}^{\left( i\right) }$ & $\left\vert I_{3,4}\left( \left\vert \Phi
_{134}^{\left( i\right) }\right\rangle \right) \right\vert $ & $\left[ \tau
_{1|3}\left( \left\vert \Phi _{134}^{\left( i\right) }\right\rangle \right) %
\right] ^{2}$ \\ \hline\hline
$A_{1}A_{3}$ & $D_{\left( A_{4}\right) _{0}}^{000}-D_{\left( A_{4}\right)
_{0}}^{001}$ & $\left( 
\begin{array}{c}
D^{0000}+D^{0001} \\ 
-D^{0010}-D^{0011}%
\end{array}%
\right) $ & $D_{\left( A_{4}\right) _{1}}^{000}-D_{\left( A_{4}\right)
_{1}}^{001}$ & $M_{A_{4}}$ & $\left\vert \left( I_{3}\right)
_{A_{2}}^{\left( new\right) }\left( \left\vert \Psi _{1234}\right\rangle
\right) \right\vert $ & $\left[ \tau _{1|3}^{(new)}\left( \rho _{123}\right) %
\right] ^{2}$ \\ \hline\hline
$A_{1}A_{4}$ & $D_{\left( A_{2}\right) _{0}\left( A_{3}\right) _{i}}^{00}$ & 
$D_{\left( A_{3}\right) _{i}}^{000}-D_{\left( A_{3}\right) _{i}}^{001}$ & $%
D_{\left( A_{2}\right) _{1}\left( A_{3}\right) _{i}}^{00}$ & $%
N_{A_{3}}^{\left( i\right) }$ & $\left\vert I_{3,4}\left( \left\vert \Phi
_{124}^{\left( i\right) }\right\rangle \right) \right\vert $ & $\left[ \tau
_{1|4}\left( \left\vert \Phi _{124}^{\left( i\right) }\right\rangle \right) %
\right] ^{2}$ \\ \hline\hline
$A_{1}A_{4}$ & $D_{\left( A_{2}\right) _{0}}^{000}-D_{\left( A_{2}\right)
_{0}}^{001}$ & $\left( 
\begin{array}{c}
D^{0000}+D^{0010} \\ 
-D^{0001}-D^{0011}%
\end{array}%
\right) $ & $D_{\left( A_{2}\right) _{1}}^{000}-D_{\left( A_{2}\right)
_{1}}^{001}$ & $M_{A_{2}}$ & $\left\vert \left( I_{3}\right)
_{A_{3}}^{\left( new\right) }\left( \left\vert \Psi _{1234}\right\rangle
\right) \right\vert $ & $\left[ \tau _{1|4}^{\left( new\right) }\left( \rho
_{134}\right) \right] ^{2}$ \\ \hline\hline
\end{tabular}%
\end{table}
Three-qubit invariants listed in the last three columns depend on two-qubit
invariants of columns two to four in the same way as $I^{(1)}$, $I^{(2)}$and 
$\tau ^{2}$ depend on $A$, $B$ and $C$, for example three-qubit invariants
in the third row of Table II read as%
\begin{equation}
N_{A_{4}}^{\left( i\right) }=\left\vert D_{\left( A_{3}\right) _{0}\left(
A_{4}\right) _{i}}^{00}\right\vert ^{2}+\frac{1}{2}\left\vert D_{\left(
A_{4}\right) _{i}}^{000}+D_{\left( A_{4}\right) _{i}}^{001}\right\vert
^{2}+\left\vert D_{\left( A_{3}\right) _{1}\left( A_{4}\right)
_{i}}^{00}\right\vert ^{2},\left( i=0,1\right)
\end{equation}%
and%
\begin{equation}
\left\vert I_{3,4}\left( \left\vert \Phi _{123}^{i}\right\rangle \right)
\right\vert =\left\vert \left( D_{\left( A_{4}\right) _{i}}^{000}+D_{\left(
A_{4}\right) _{i}}^{001}\right) ^{2}-4D_{\left( A_{3}\right) _{0}\left(
A_{4}\right) _{i}}^{00}D_{\left( A_{3}\right) _{1}\left( A_{4}\right)
_{i}}^{00}\right\vert ,
\end{equation}%
and satisfy the inequality%
\begin{equation}
4N_{A_{4}}^{\left( i\right) }\geq \left[ \tau _{1|2}\left( \left\vert \Phi
_{123}^{\left( i\right) }\right\rangle \right) \right] ^{2}+\frac{1}{2}%
\left\vert 4I_{3,4}\left( \left\vert \Phi _{123}^{\left( i\right)
}\right\rangle \right) \right\vert .  \label{N4_sum}
\end{equation}%
Here $\left\vert \Phi _{123}^{\left( i\right) }\right\rangle $ is the
un-normalized state defined through $\left\vert \Psi _{1234}\right\rangle
=\sum_{i_{4}=0,1}\left\vert \Phi _{123}^{\left( i_{4}\right) }\right\rangle
\left\vert i_{4}\right\rangle $.

Upper bound on two-tangle calculated by using the method of ref. \cite%
{shar16} shows that%
\begin{equation}
\sum\limits_{i=0,1}\left[ \tau _{1|2}\left( \left\vert \Phi _{123}^{\left(
i\right) }\right\rangle \right) \right] ^{2}\geq \left[ \tau
_{1|2}^{up}\left( \rho _{12}\right) \right] ^{2}\geq \left[ \tau
_{1|2}\left( \rho _{12}\right) \right] ^{2}.  \label{uptwo}
\end{equation}%
Similarly the upper bounds on $\tau _{1|2|3}\left( \rho _{123}\right) $ for
the nine families of four-qubit states, calculated in ref. \cite{shar216}
satisfy the condition 
\begin{equation}
\sum\limits_{i=0,1}\left\vert 4I_{3,4}\left( \left\vert \Phi _{123}^{\left(
i\right) }\right\rangle \right) \right\vert \geq \tau _{1|2|3}^{up}\left(
\rho _{123}\right) \geq \tau _{1|2|3}\left( \rho _{123}\right) \text{.}
\label{upthree}
\end{equation}%
Combining the conditions of Eqs. (\ref{uptwo}) and (\ref{upthree}), with
inequality of Eq. (\ref{N4_sum}), the sum of two-tangle and three-tangle
satisfies the inequality 
\begin{equation}
4\sum\limits_{i=0,1}N_{A_{4}}^{\left( i\right) }\geq \left[ \tau
_{1|2}\left( \rho _{12}\right) \right] ^{2}+\frac{1}{2}\tau _{1|2|3}\left(
\rho _{123}\right) .  \label{N4}
\end{equation}%
On $\rho _{124}=\sum\limits_{i}\left\vert \Phi _{124}^{i}\right\rangle
\left\langle \Phi _{124}^{i}\right\vert $, new two-qubit invariant (Eq. (\ref%
{1pnew})) is defined as%
\begin{equation*}
\tau _{1|2}^{\left( new\right) }\left( \rho _{124}\right) =2\min_{\left\{
\left\vert \Phi _{124}^{\left( i\right) }\right\rangle \right\}
}\sum\limits_{i}\left\vert T_{12}\left( \left\vert \Phi _{124}^{\left(
i\right) }\right\rangle \right) \right\vert \text{.}
\end{equation*}%
where $T_{12}\left( \left\vert \Phi _{124}^{\left( i\right) }\right\rangle
\right) =D_{\left( A_{3}\right) _{i}}^{000}\left( \left\vert \Phi
_{124}^{\left( i\right) }\right\rangle \right) +D_{\left( A_{3}\right)
_{i}}^{001}\left( \left\vert \Phi _{124}^{\left( i\right) }\right\rangle
\right) $. New three-qubit tangle on a pure state is defined as $\tau
_{1|2|3}^{\left( new\right) }\left( \left\vert \Psi _{1234}\right\rangle
\right) =4\left\vert \left( I_{3}\right) _{A_{4}}^{\left( new\right) }\left(
\left\vert \Psi _{1234}\right\rangle \right) \right\vert $, where%
\begin{eqnarray*}
\left( I_{3}\right) _{A_{4}}^{\left( new\right) }\left( \left\vert \Psi
_{1234}\right\rangle \right) &=&\left(
D^{0000}+D^{0010}+D^{0001}+D^{0011}\right) ^{2} \\
&&-4\left( D_{\left( A_{3}\right) _{0}}^{000}+D_{\left( A_{3}\right)
_{0}}^{001}\right) \left( D_{\left( A_{3}\right) _{1}}^{000}+D_{\left(
A_{3}\right) _{1}}^{001}\right) .
\end{eqnarray*}%
The invariants $M_{A_{3}}$, $\tau _{1|2|3}^{\left( new\right) }\left(
\left\vert \Psi _{1234}\right\rangle \right) $ and $\left[ \tau
_{1|2}^{\left( new\right) }\left( \rho _{124}\right) \right] ^{2}$ satisfy
the inequality (analogous to Eq. (\ref{iineq})),%
\begin{equation}
4M_{A_{3}}-\frac{1}{2}\tau _{1|2|3}^{\left( new\right) }\left( \left\vert
\Psi _{1234}\right\rangle \right) \geq \left[ \tau _{1|2}^{\left( new\right)
}\left( \rho _{124}\right) \right] ^{2}.  \label{M123}
\end{equation}

Using a similar argument, three-qubit invariants listed in lines 5 and 6 of
Table II satisfy the inequalities%
\begin{equation}
4\sum\limits_{i}N_{A_{2}}^{\left( i\right) }\geq \left[ \tau _{1|3}\left(
\rho _{13}\right) \right] ^{2}+\frac{1}{2}\tau _{1|3|4}\left( \rho
_{134}\right) \text{,}  \label{N2}
\end{equation}%
and%
\begin{equation}
4M_{A_{4}}-\frac{1}{2}\tau _{1|3|4}^{\left( new\right) }\left( \left\vert
\Psi _{1234}\right\rangle \right) \geq \left[ \tau _{1|3}^{\left( new\right)
}\left( \rho _{123}\right) \right] ^{2},  \label{M134}
\end{equation}%
where three-tangle defined on pure four-qubit state reads as $\tau
_{1|3|4}^{\left( new\right) }\left( \left\vert \Psi _{1234}\right\rangle
\right) =4\left\vert \left( I_{3}\right) _{A_{2}}^{\left( new\right) }\left(
\left\vert \Psi _{1234}\right\rangle \right) \right\vert $, and%
\begin{equation*}
\tau _{1|3}^{\left( new\right) }\left( \rho _{123}\right) =2\min_{\left\{
\left\vert \Phi _{123}^{\left( i\right) }\right\rangle \right\}
}\sum\limits_{i}\left\vert T_{13}\left( \left\vert \Phi _{123}^{\left(
i\right) }\right\rangle \right) \right\vert .
\end{equation*}

Using invariants of local unitaries on qubits $A_{1}$ and $A_{4}$, and
definitions given in lines 7 and 8 of Table II, we obtain the inequalities%
\begin{equation}
4\sum\limits_{i}N_{A_{3}}^{\left( i\right) }\geq \left[ \tau _{1|4}\left(
\rho _{14}\right) \right] ^{2}+\frac{1}{2}\tau _{1|2|4}\left( \rho
_{124}\right)  \label{N3}
\end{equation}%
and%
\begin{equation}
4M_{A_{2}}-\frac{1}{2}\tau _{1|2|4}^{\left( new\right) }\left( \Psi
_{1234}\right) \geq \left[ \tau _{1|4}^{\left( new\right) }\left( \rho
_{134}\right) \right] ^{2},  \label{M124}
\end{equation}%
where new three-tangle reads as $\tau _{1|2|4}^{\left( new\right) }\left(
\Psi _{1234}\right) =4\left\vert \left( I_{3}\right) _{A_{3}}^{\left(
new\right) }\left( \left\vert \Psi _{1234}\right\rangle \right) \right\vert $%
, and 
\begin{equation*}
\tau _{1|4}^{\left( new\right) }\left( \rho _{134}\right) =2\min_{\left\{
\left\vert \Phi _{134}^{\left( i\right) }\right\rangle \right\}
}\sum\limits_{i}\left\vert D_{\left( A_{2}\right) _{i}}^{000}\left(
\left\vert \Phi _{134}^{\left( i\right) }\right\rangle \right) -D_{\left(
A_{2}\right) _{i}}^{001}\left( \left\vert \Phi _{134}^{\left( i\right)
}\right\rangle \right) \right\vert .
\end{equation*}

The relations between two-tangles, three-tangles and three-qubit invariants
listed in column (5) in Table II (Eqs. (\ref{N4}-\ref{M124})) are important
to obtain the monogamy inequality satisfied by one-tangle.

\section{Monogamy of four-qubit entanglement}

To obtain the relation between tangles of reduced states and one-tangle of
the focus qubit, firstly, we identify the three-qubit invariant combinations
of two-qubit invariants in Eq. (\ref{tanglea1}). It is found that a
four-qubit invariant of degree two, which is defined only on the pure state,
is also needed. Genuine four-tangle $\tau _{1|2|3|4}\left( \left\vert \Psi
_{1234}\right\rangle \right) $ (Eq. (\ref{genuine}) appendix \ref{A2}),
defined in refs. \cite{shar14,shar16} is a degree-eight function of state
coefficients. However, the degree-two four-qubit invariant which is equal to
invariant H of refs. \cite{shar10,luqu03}, is known to have the form,%
\begin{equation}
I_{4,2}\left( \left\vert \Psi _{1234}\right\rangle \right)
=D^{0000}-D^{0010}-D^{0001}+D^{0011}.  \label{fort0}
\end{equation}%
Four-tangle defined as $\tau _{1|2|3|4}^{(0)}\left( \left\vert \Psi
_{1234}\right\rangle \right) =2\left\vert I_{4,2}\left( \left\vert \Psi
_{1234}\right\rangle \right) \right\vert $, is non zero on a GHZ state and
vanishes on W-like states of four qubits. However, since $\tau
_{1|2|3|4}^{(0)}\left( \left\vert \Psi _{1234}\right\rangle \right) $ fails
to vanish on product of entangled states of two qubits, it is not a measure
of genuine four-way entanglement. By direct substitution, one-tangle of Eq. (%
\ref{tanglea1}) can be rewritten in terms of three-qubit invariants\ listed
in column five of Table II and square of four-qubit invariant $\tau
_{1|2|3|4}^{(0)}\left( \left\vert \Psi _{1234}\right\rangle \right) $ that is%
\begin{equation}
\tau _{1|234}=4\sum\limits_{q=2}^{4}\sum\limits_{i=0}^{1}N_{A_{q}}^{\left(
i\right) }+2\sum\limits_{q=2}^{4}M_{A_{q}}+\frac{1}{4}\left[ \tau
_{1|2|3|4}^{(0)}\left( \left\vert \Psi _{1234}\right\rangle \right) \right]
^{2}.  \label{tangle2}
\end{equation}

By making use of the inequalities given by Eqs. (\ref{N4}-\ref{M124}),
one-tangle of Eq. (\ref{tangle2}) is found to satisfy the constraint%
\begin{eqnarray}
\tau _{1|234}\left( \left\vert \Psi _{1234}\right\rangle \right) &\geq
&\sum\limits_{p=2}^{4}\left[ \tau _{1|p}\left( \rho _{1p}\right) \right]
^{2}+\frac{1}{2}\sum\limits_{\substack{ \left( p,q\right) =2  \\ q>p}}%
^{4}\tau _{1|p|q}\left( \rho _{1pq}\right)  \notag \\
&&+\frac{1}{2}\left[ \tau _{1|2}^{\left( new\right) }\left( \rho
_{123}\right) \right] ^{2}+\frac{1}{2}\left[ \tau _{1|3}^{\left( new\right)
}\left( \rho _{134}\right) \right] ^{2}+\frac{1}{2}\left[ \tau
_{1|4}^{\left( new\right) }\left( \rho _{124}\right) \right] ^{2}.
\label{mono}
\end{eqnarray}%
This is the monogamy relation that governs the distribution of quantum
entanglement in subsystems of a four-qubit state, when qubit $A_{1}$ is the
focus qubit. Similar inequalities can be written down for other possible
choices of focus qubit.

Parameter $\Delta $, which quantifies the residual correlations not
accounted for by entanglement of reduced states, is defined as the
difference between the entanglement of focus qubit $A_{1}$ with the rest of
the system and contributions from sum of two-way and three-way tangles of
focus qubit, that is%
\begin{eqnarray}
\Delta &=&\tau _{1|234}\left( \left\vert \Psi _{1234}\right\rangle \right)
-\sum\limits_{p=2}^{4}\left[ \tau _{1|p}\left( \rho _{1p}\right) \right]
^{2}-\frac{1}{2}\sum\limits_{\substack{ \left( p,q\right) =2  \\ q>p}}%
^{4}\tau _{1|p|q}\left( \rho _{1pq}\right)  \notag \\
&&-\frac{1}{2}\left[ \tau _{1|2}^{\left( new\right) }\left( \rho
_{123}\right) \right] ^{2}-\frac{1}{2}\left[ \tau _{1|3}^{\left( new\right)
}\left( \rho _{134}\right) \right] ^{2}-\frac{1}{2}\left[ \tau
_{1|4}^{\left( new\right) }\left( \rho _{124}\right) \right] ^{2}.
\label{delmono}
\end{eqnarray}%
All contributions to $\Delta $ are invariant with respect to local unitaries
on any one of the four qubits. A possible lower bound on $\Delta $ which,
partially, accounts for residual four-qubit correlations depends on the sum
of three-qubit invariants $\tau _{1|p|q}^{\left( new\right) }\left(
\left\vert \Psi _{1234}\right\rangle \right) $, four-qubit invariant $\tau
_{1|2|3|4}^{(0)}\left( \left\vert \Psi _{1234}\right\rangle \right) $ and
genuine four-tangle $\tau _{1|2|3|4}\left( \left\vert \Psi
_{1234}\right\rangle \right) $. It will be interesting to write residual
four-qubit correlations as a permutationally invariant combination of
four-qubit invariants defined on the state $\left\vert \Psi
_{1234}\right\rangle $ such that all possible modes of four-way entanglement
are accounted for.

\section{Lower bound on residual four-qubit correlations}

In this section, we examine the known four-qubit invariants to identify the
possible candidates to represent four-way correlations and construct a lower
bound on residual correlations. Transformation equation of determinant of a
two-way negativity font due to the action of local unitaries on the two
remaining qubits, yields a two-variable polynomial of degree four.
Invariants of the polynomials corresponding to qubit pairs $A_{1}A_{2}$, $%
A_{1}A_{3}$, and $A_{1}A_{4}$ are four-qubit invariants. For instance,
four-qubit invariants corresponding to $U^{3}U^{4}D_{\left( A_{3}\right)
_{0}\left( A_{4}\right) _{0}}^{00}=0$, are $N_{A_{4}}^{\left( 0\right) }+%
\frac{1}{2}M_{A_{3}}+N_{A_{4}}^{\left( 1\right) }$ and%
\begin{eqnarray}
J^{A_{1}A_{2}} &=&\left( I_{3}\right) _{A_{4}}^{\left( new\right) }\left(
\left\vert \Psi _{1234}\right\rangle \right) -4\left( D_{\left( A_{4}\right)
_{0}}^{000}+D_{\left( A_{4}\right) _{0}}^{001}\right) \left( D_{\left(
A_{4}\right) _{1}}^{000}+D_{\left( A_{4}\right) _{1}}^{001}\right)  \notag \\
&&+8\left( D_{\left( A_{3}\right) _{1}\left( A_{4}\right)
_{0}}^{00}D_{\left( A_{3}\right) _{0}\left( A_{4}\right)
_{1}}^{00}+D_{\left( A_{3}\right) _{0}\left( A_{4}\right)
_{0}}^{00}D_{\left( A_{3}\right) _{1}\left( A_{4}\right) _{1}}^{00}\right) .
\label{i1234}
\end{eqnarray}%
Notice that $\left\vert 4J^{A_{1}A_{2}}\right\vert $ contains contribution
from $\tau _{1|2|3}^{\left( new\right) }\left( \left\vert \Psi
_{1234}\right\rangle \right) $.

Similarly, four-qubit invariant that quantifies the entanglement of $A_{1}$
and $A_{3}$ in $N_{A_{2}}^{\left( 0\right) }+\frac{1}{2}%
M_{A_{4}}+N_{A_{2}}^{\left( 1\right) }$ is obtained from%
\begin{eqnarray}
J^{A_{1}A_{3}} &=&\left( I_{3}\right) _{A_{2}}^{\left( new\right) }\left(
\left\vert \Psi _{1234}\right\rangle \right) -4\left( D_{\left( A_{2}\right)
_{0}}^{000}+D_{\left( A_{2}\right) _{0}}^{001}\right) \left( D_{\left(
A_{2}\right) _{1}}^{000}+D_{\left( A_{2}\right) _{1}}^{001}\right)  \notag \\
&&+8\left( D_{\left( A_{2}\right) _{1}\left( A_{4}\right)
_{0}}^{00}D_{\left( A_{2}\right) _{0}\left( A_{4}\right)
_{1}}^{00}+D_{\left( A_{2}\right) _{0}\left( A_{4}\right)
_{0}}^{00}D_{\left( A_{2}\right) _{1}\left( A_{4}\right) _{1}}^{00}\right)
\label{1342}
\end{eqnarray}%
while the sum $N_{A_{3}}^{\left( 0\right) }+\frac{1}{2}M_{A_{2}}+N_{A_{3}}^{%
\left( 1\right) }$ contains correlations between qubits $A_{1}$ and $A_{4}$
which depend on the four-qubit invariant that reads as%
\begin{eqnarray}
J^{A_{1}A_{4}} &=&\left( I_{3}\right) _{A_{3}}^{\left( new\right) }\left(
\left\vert \Psi _{1234}\right\rangle \right) -4\left( D_{\left( A_{3}\right)
_{0}}^{000}-D_{\left( A_{3}\right) _{0}}^{001}\right) \left( D_{\left(
A_{3}\right) _{1}}^{000}-D_{\left( A_{3}\right) _{1}}^{001}\right)  \notag \\
&&+8\left( D_{\left( A_{2}\right) _{0}\left( A_{3}\right)
_{1}}^{00}D_{\left( A_{2}\right) _{1}\left( A_{3}\right)
_{0}}^{00}+D_{\left( A_{2}\right) _{0}\left( A_{3}\right)
_{0}}^{00}D_{\left( A_{2}\right) _{1}\left( A_{3}\right) _{1}}^{00}\right) .
\label{i1423}
\end{eqnarray}%
It is easily verified \cite{shar10} that $J^{A_{1}A_{i}}$ ($i=2,3,4$) are
not independent invariants but satisfy the constraint%
\begin{equation*}
4\left\vert J^{A_{1}A_{2}}+J^{A_{1}A_{3}}+J^{A_{1}A_{4}}\right\vert =3\left[
\tau _{1|2|3|4}^{(0)}\left( \left\vert \Psi _{1234}\right\rangle \right) %
\right] ^{2}.
\end{equation*}%
By construction, $\beta ^{A_{1}A_{i}}=\frac{4}{3}\left\vert
J^{A_{1}A_{i}}\right\vert $ is the entanglement of qubit pair $A_{1}A_{i}$
in the four-qubit state due to two-way, three-way and four-way correlations.
It is easily verified that $\left\vert J^{A_{1}A_{2}}\right\vert =\left\vert
J^{A_{3}A_{4}}\right\vert $, $\left\vert J^{A_{1}A_{3}}\right\vert
=\left\vert J^{A_{2}A_{4}}\right\vert $, and $\left\vert
J^{A_{1}A_{4}}\right\vert =\left\vert J^{A_{2}A_{3}}\right\vert $, as such, $%
\sum_{i\neq p}\beta ^{A_{p}A_{i}}$ does not depend on the choice of focus
qubit. A four-qubit state has four-way correlations if and only if at least
two of the three $\beta ^{A_{1}A_{i}}$ ($i=2,3,4$) are non-zero. Recalling
that one-tangle has a contribution from four-qubit invariant $\tau
_{1|2|3|4}^{(0)}\left( \left\vert \Psi _{1234}\right\rangle \right) $, and
genuine four-tangle $\tau _{1|2|3|4}$ (Eq. (\ref{genuine})), a lower bound
on residual four-way correlations for the set of state, satisfying $%
\sum_{i,j=2;(i<j)}^{4}\beta ^{A_{1}A_{i}}\beta ^{A_{1}A_{j}}\neq 0$, can be
written as%
\begin{equation}
\delta (\left\vert \Psi _{1234}\right\rangle )=\frac{1}{4}%
\sum_{i=2}^{4}\beta ^{A_{1}A_{i}}+\frac{1}{4}\left[ \tau
_{1|2|3|4}^{(0)}\left( \left\vert \Psi _{1234}\right\rangle \right) \right]
^{2}+\frac{1}{2}\sqrt{\tau _{1|2|3|4}}.  \label{deltatheory}
\end{equation}
All contributions to $\delta \left( \left\vert \Psi _{1234}\right\rangle
\right) $ are invariant with respect to local unitaries on any one of the
four qubits as well as the choice of focus qubit.

On a four-qubit GHZ state 
\begin{equation}
\left\vert GHZ_{4}\right\rangle =\frac{1}{\sqrt{2}}\left( \left\vert
0000\right\rangle +\left\vert 1111\right\rangle \right) ,
\end{equation}%
which is known to have only four-qubit correlations 
\begin{equation*}
\tau _{1|234}=\tau _{1|2|3|4}=\left[ \tau _{1|2|3|4}^{(0)}\left( \left\vert
\Psi _{1234}\right\rangle \right) \right] ^{2}=1,
\end{equation*}%
while $\beta ^{A_{1}A_{i}}=\frac{1}{3}$ ($i=2,3,4$). Therefore, as expected $%
\tau _{1|234}=\delta (\left\vert \Psi _{1234}\right\rangle )$. While on the
state%
\begin{equation}
\left\vert \Phi \right\rangle =\frac{1}{2}\left( \left\vert
1111\right\rangle +\left\vert 1100\right\rangle +\left\vert
0010\right\rangle +\left\vert 0001\right\rangle \right) ,
\end{equation}%
with no three-tangles and two-tangles, we have $\beta ^{A_{1}A_{2}}=\frac{2}{%
3}$, $\beta ^{A_{1}A_{3}}=\beta ^{A_{1}A_{4}}=\frac{1}{3}$ and%
\begin{equation*}
\tau _{1|234}=\tau _{1|2|3|4}=1;\tau _{1|2|3|4}^{(0)}\left( \left\vert \Psi
_{1234}\right\rangle \right) =0,
\end{equation*}%
therefore $\delta \left( \left\vert \Phi \right\rangle \right) =\frac{5}{6}$%
. In this case not all four-way correlations are accounted for by $\delta
\left( \left\vert \Phi \right\rangle \right) $.

\section{ States violating general Monogamy inequality}

As mentioned in ref. \cite{regu14} a natural extension of CKW inequality to
four-qubit states reads as%
\begin{equation}
\mathcal{\tau }_{1|234}\left( \left\vert \Psi _{1234}\right\rangle \right)
\geq \sum\limits_{p=2}^{4}\left[ \tau _{1|p}\left( \rho _{1p}\right) \right]
^{2}+\sum\limits_{\substack{ \left( p,q\right) =2 \\ q>p}}^{4}\tau
_{1|p|q}\left( \rho _{1pq}\right) .  \label{mono1}
\end{equation}%
Regula et al. \cite{regu14} have analysed arbitrary pure states $\left\vert
\Psi _{1234}\right\rangle $ of four-qubit systems and shown that a subset of
these states violates the inequality of Eq. (\ref{mono1}). Based on
numerical evidence, the authors provide a conjectured monogamy inequality.
For the case of four-qubits with $A_{1}$ as focus qubit, the monogamy
inequality of Eq. (9) in ref. \cite{regu14} reads as%
\begin{eqnarray}
\mathcal{\tau }_{1|234}\left( \left\vert \Psi _{1234}\right\rangle \right) 
&\geq &\left[ \mathcal{\tau }_{1|2}\left( \rho _{12}\right) \right] ^{2}+%
\left[ \mathcal{\tau }_{1|3}\left( \rho _{13}\right) \right] ^{2}+\left[ 
\mathcal{\tau }_{1|4}\left( \rho _{14}\right) \right] ^{2}  \notag \\
&&+\left[ \mathcal{\tau }_{1|2|3}\left( \rho _{123}\right) \right] ^{\frac{3%
}{2}}+\left[ \mathcal{\tau }_{1|2|4}\left( \rho _{124}\right) \right] ^{%
\frac{3}{2}}+\left[ \mathcal{\tau }_{1|3|4}\left( \rho _{134}\right) \right]
^{\frac{3}{2}}.  \label{mono2}
\end{eqnarray}%
Here three tangles are raised to the power $\frac{3}{2}$, so that the
\textquotedblleft residual four tangle\textquotedblright\ may not become
negative. We denote the residual correlations, calculated from inequality of
Eq. (\ref{mono1}) by $\Delta _{1}$ and that from inequality proposed in ref. 
\cite{regu14} (Eq. (\ref{mono2})) by $\Delta _{2}$. In a more recent article 
\cite{regu16}, it has been clarified that the states leading to violations
of the strong monogamy inequality belong to the degenerate subclasses (with $%
a=c$ or $b=c)$ of $\left\vert G_{abc}^{\left( 2\right) }\right\rangle $ \cite%
{vers02} defined as%
\begin{align}
\left\vert G_{abc}^{\left( 2\right) }\right\rangle & =\frac{a+b}{2}\left(
\left\vert 0000\right\rangle +\left\vert 1111\right\rangle \right) +\frac{a-b%
}{2}\left( \left\vert 0011\right\rangle +\left\vert 1100\right\rangle
\right)   \notag \\
& +c\left( \left\vert 0101\right\rangle +\left\vert 1010\right\rangle
\right) +\left\vert 0110\right\rangle .
\end{align}%
For the choice $b=c=ia$ with $a\geq 0$, we obtain the class of single
parameter states $\left\vert G_{a,ia,ia}^{\left( 2\right) }\right\rangle $.
It is found that with qubit $A_{1}$ as the focus qubit, one-tangle takes the
value%
\begin{equation}
\tau _{1|234}=\frac{8a^{2}+16a^{4}}{\left( 4a^{2}+1\right) ^{2}},
\end{equation}%
while $\tau _{1|2|3|4}^{(0)}=\frac{2a^{2}}{\left( 4a^{2}+1\right) }$ and
genuine four-tangle $\tau _{1|2|3|4}=0$. Three tangles for the states read
as $\tau _{1|2|3}=\tau _{1|2|4}=\tau _{1|3|4}=\frac{8a^{3}}{\left(
4a^{2}+1\right) ^{2}}.$ All new two-tangles take value zero. Two-qubit
states obtained from $\left\vert G_{a,ia,ia}^{\left( 2\right) }\right\rangle 
$ after tracing out a pair of qubits are $X$ states. Two-tangles for these
states were calculated numerically.

\begin{figure}[tbph]
\centering \includegraphics[width=5.0in]{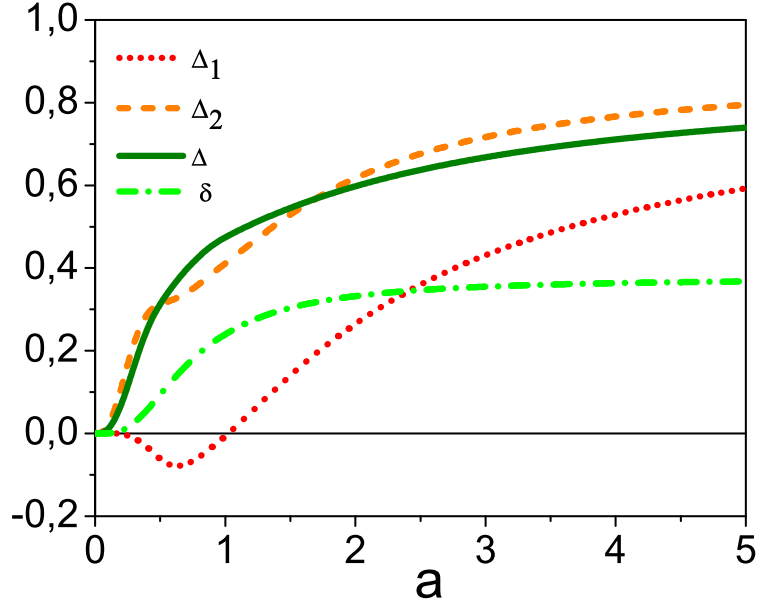} 
\caption{ Plot of $\Delta $ ( Olive green line-Solid) from our monogamy
relation Eq. (\protect\ref{delmono}), $\Delta _{1}$(red line-Dot) from
monogamy inequality of Eq. (\protect\ref{mono1}) and $\Delta _{2}$ (orange
line-Dash) using monogamy conjecture of ref. \protect\cite{regu14} Eq. (%
\protect\ref{mono2}), for the states $\left\vert G_{a,ia,ia}^{\left(
2\right) }\right\rangle $ ( $0\leq a\leq 5$). Figure also displays a plot of
our lower bound on residual four-qubit correlations $\protect\delta \left(
\left\vert G_{a,ia,ia}^{\left( 2\right) }\right\rangle \right) $ (green
line-Dash Dot). }
\label{fig1}
\end{figure}

Figure I displays the residual four-qubit correlations, $\Delta $ (Olive
green line-Solid) from our monogamy relation (Eq. (\ref{delmono})), $\Delta
_{1}$(red line-Dot) obtained by using a generalization of CKW inequality
(Eq. (\ref{mono1})), as well as $\Delta _{2}$ (orange line-Dash) from
monogamy conjecture of ref. \cite{regu14} (Eq. (\ref{mono2})), for the
states $\left\vert G_{a,ia,ia}^{\left( 2\right) }\right\rangle $ ( $0\leq
a\leq 5$). As expected $\Delta _{1}$ becomes negative for a certain range of
values of $a$, while $\Delta $ and $\Delta _{2}$ remain positive for the set
of states being considered. Since three-tangle varies from zero to one, and $%
\left( \mathcal{\tau }_{1|p|q}\right) ^{\frac{3}{2}}>\frac{1}{2}\mathcal{%
\tau }_{1|p|q}$ for $\mathcal{\tau }_{1|p|q}>0.25$, it is obvious that the
conjecture of ref. \cite{regu14} either overestimates or underestimates the
four-way correlations in classes of four-qubit states with finite three
tangles, but no contribution from new two-tangles. The lower bound on
residual four-qubit correlations is found to be (Eq. (\ref{deltatheory}))%
\begin{equation}
\delta \left( \left\vert G_{a,ia,ia}^{\left( 2\right) }\right\rangle \right)
=\frac{6a^{4}}{\left( 4a^{2}+1\right) ^{2}}\text{.}
\end{equation}%
Figure I also shows a plot (green line-Dash Dot) of $\delta \left(
\left\vert G_{a,ia,ia}^{\left( 2\right) }\right\rangle \right) $.

\section{Conclusions}

Monogamy inequality of Eq. (\ref{mono}), obtained analytically by expressing
the tangle in terms of two-qubit and three-qubit invariants, is satisfied by
all classes of four-qubit states. In particular, it is satisfied on the set
of states $\left\vert G_{a,ia,ia}^{(2)}\right\rangle $ that violate the
entanglement monogamy relation obtained by a generalization of CKW
inequality. The residual four-qubit correlations obtained by subtracting
two-tangles and three-tangles from one-tangle of focus qubit represent
contributions from all possible four-qubit entanglement modes. Unlike the
case of three-qubits, where a single degree-four invariant quantifies
three-way correlations in a pure state, a combination of four-qubit
invariants is needed to quantify all possible modes of four-qubit pure state
entanglement. Lower bound constructed from four-qubit invariants of degree $%
2 $, $4$, and $8$ is permutationally invariant and partially accounts for
residual four-way correlations. To close the question, it will be
interesting to find a single four-qubit invariant or write a combination of
known four-qubit invariants which accounts for the residual correlations in
all four-qubit pure states.

On the states $\left\vert G_{a,ia,ia}^{(2)}\right\rangle $, value of genuine
four-tangle is found to be zero, while the residual entanglement is non
zero. Since on $\left\vert G_{a,ia,ia}^{(2)}\right\rangle $ all three
tangles are finite for $a\neq 0,$ four qubits are entangled through
three-way correlations. Four-qubit states may, likewise, be entangled due to
combination of three-way and two-way correlations or only pairwise
correlations. Consider the subset of pure states in which four-qubit
entanglement arises due to three-way correlations, while genuine four-way
correlations are absent. The monogamy inequality conjecture for four qubits
in which quantifiers of three-way entanglement (three tangles) should be
raised to the power $\frac{3}{2}$ \cite{regu14} does not estimate the
residual correlations, correctly, for these states. A simple example is the
state%
\begin{equation*}
\left\vert \Psi \right\rangle =\frac{1}{2}\left( \left\vert
0000\right\rangle +\left\vert 0101\right\rangle +\left\vert
1000\right\rangle +\left\vert 1110\right\rangle \right) ,
\end{equation*}%
for which $\Delta =0$, while Eq. (\ref{mono2}) predicts $\Delta _{2}=\frac{3%
}{8}$. In general, the approach of ref. \cite{regu14}, where an $n-$tangle
is raised to an arbitrary power to account for $n-$way correlations in
one-tangle, is not likely to account for $n-$way correlations, correctly,
for all the states. A simple calculation, on the same lines as for the case
of four qubits, shows that as more qubits are added the contribution to
one-tangle from three-tangles defined on three-qubit pure states is always
multiplied by a factor of $\frac{1}{2}$. In addition, one-tangle may have a
contribution from new-two tangles. For a given value of three-tangle,
contribution from corresponding new two-tangles does not have a simple
relationship with the three-tangle. In a multi-qubit state, the contribution
of degree four $n$-tangle $\left( \tau _{1|2|...|n}\right) $ to degree four
one-tangle will be multiplied by a factor of $\left( \frac{1}{2}\right)
^{n-2}$. Our approach paves the way to understanding scaling of entanglement
distribution as qubits are added to obtain larger multiqubit quantum systems.

Financial support from Universidade Estadual de Londrina PR, Brazil is
acknowledged.

\appendix

\section{Upper bound on a function of complex numbers}

\label{A1}

In this section, we obtain mathematical relations used in section \ref%
{tangles4q} to define necessary invariants on four-qubit and three-qubit
states. Let $A(x)$ , $B\left( x\right) $\ and $C(x)$ satisfy the set of
equations%
\begin{equation}
A(x)=\frac{1}{1+\left\vert x\right\vert ^{2}}\left( A-x^{\ast }B+\left(
x^{\ast }\right) ^{2}C\right) ,  \label{ax}
\end{equation}%
\begin{equation}
B\left( x\right) =\frac{1}{1+\left\vert x\right\vert ^{2}}\left( B\left(
1-\left\vert x\right\vert ^{2}\right) -2x^{\ast }C+2xA\right) ,  \label{bx}
\end{equation}%
\begin{equation}
C(x)=\frac{1}{1+\left\vert x\right\vert ^{2}}\left( C+xB+x^{2}A\right) ,
\label{cx}
\end{equation}%
where $A$, $B$, and $C$ are complex numbers. One can verify that 
\begin{equation}
I^{(1)}=\left\vert A\right\vert ^{2}+\frac{1}{2}\left\vert B\right\vert
^{2}+\left\vert C\right\vert ^{2}=\left\vert A(x)\right\vert ^{2}+\frac{1}{2}%
\left\vert B(x)\right\vert ^{2}+\left\vert C(x)\right\vert ^{2}.  \label{i1}
\end{equation}%
Consider a function of complex variable $x$ defined as%
\begin{equation}
\tau =2\min_{x}\left( \left\vert A(x)\right\vert +\left\vert C(x)\right\vert
\right) =\min_{x}I(x),
\end{equation}%
Value of $x$ for which $A(x)=0$ is%
\begin{equation*}
x_{0}^{\ast }=\frac{B}{2C}\pm \frac{1}{2C}\sqrt{B^{2}-4AC}.
\end{equation*}%
The discriminant $B^{2}-4AC$ is, obviously, an invariant, so we define $%
I^{(2)}=\left\vert B^{2}-4AC\right\vert $. Substituting the value of $x_{0}$
in Eq. (\ref{cx}), we obtain%
\begin{equation}
I(x_{0})=2\left\vert C(x_{0})\right\vert =2\sqrt{I^{(1)}-\frac{1}{2}I^{(2)}}%
;I^{(1)}\geq \frac{1}{2}I^{(2)}  \label{iup}
\end{equation}%
From the definition of $\tau $, $I(x_{0})$ is an upper bound on $I$ that is 
\begin{equation}
\tau ^{2}\leq 4I^{(1)}-2I^{(2)}.  \label{iineq}
\end{equation}

\section{Genuine four-tangle}

\label{A2}

Three-qubit and four-qubit invariants relevant to quantifying three-way and
genuine four-way entanglement of state (\ref{4state}) had been constructed
in ref. \cite{shar16}. Degree four invariants of interest for the triple $%
A_{1}A_{3}A_{4}$ in state $\left\vert \Psi _{1234}\right\rangle $ comprise a
set denoted by $\left\{ \left( I_{3,4}\right) _{A_{2}}^{4-m,m}:m=0\text{ to }%
4\right\} $. The elements in the set $\left\{ \left( I_{3,4}\right)
_{A_{2}}^{4-m,m}:m=0\text{ to }4\right\} $ are invariant with respect to
local unitaries on qubits $A_{1}$, $A_{3}$, and $A_{4}$. The three-qubit
invariants for $A_{1}A_{3}A_{4}$ in terms of two-qubit invariants for the
pair $A_{1}A_{3}$ read as:%
\begin{equation}
\left( I_{3}\right) _{A_{2}}^{4,0}=\left( D_{\left( A_{2}\right)
_{0}}^{000}+D_{\left( A_{2}\right) _{0}}^{001}\right) ^{2}-4D_{\left(
A_{2}\right) _{0}\left( A_{4}\right) _{0}}^{00}D_{\left( A_{2}\right)
_{0}\left( A_{4}\right) _{1}}^{00},
\end{equation}%
\begin{eqnarray}
\left( I_{3}\right) _{A_{2}}^{3,1} &=&\frac{1}{2}\left( D_{\left(
A_{2}\right) _{0}}^{000}+D_{\left( A_{2}\right) _{0}}^{001}\right) \left(
D^{0000}+D^{0001}-D^{0010}-D^{0011}\right)  \notag \\
&&-\left[ D_{\left( A_{2}\right) _{0}\left( A_{4}\right) _{1}}^{00}\left(
D_{\left( A_{4}\right) _{0}}^{000}-D_{\left( A_{4}\right) _{0}}^{001}\right)
+D_{\left( A_{3}\right) _{0}\left( A_{4}\right) _{0}}^{00}\left( D_{\left(
A_{4}\right) _{1}}^{000}-D_{\left( A_{4}\right) _{1}}^{001}\right) \right] ,
\end{eqnarray}

\begin{eqnarray}
\left( I_{3}\right) _{A_{2}}^{2,2} &=&\frac{1}{6}\left(
D^{0000}+D^{0001}-D^{0010}-D^{0011}\right) ^{2}  \notag \\
&&-\frac{2}{3}\left( D_{\left( A_{4}\right) _{0}}^{000}-D_{\left(
A_{4}\right) _{0}}^{001}\right) \left( D_{\left( A_{4}\right)
_{1}}^{000}-D_{\left( A_{4}\right) _{1}}^{001}\right)  \notag \\
&&+\frac{1}{3}\left( D_{\left( A_{2}\right) _{0}}^{000}+D_{\left(
A_{2}\right) _{0}}^{001}\right) \left( D_{\left( A_{2}\right)
_{1}}^{000}+D_{\left( A_{2}\right) _{1}}^{001}\right)  \notag \\
&&-\frac{2}{3}\left( D_{\left( A_{2}\right) _{0}\left( A_{4}\right)
_{1}}^{00}D_{\left( A_{2}\right) _{1}\left( A_{4}\right)
_{0}}^{00}+D_{\left( A_{2}\right) _{0}\left( A_{4}\right)
_{0}}^{00}D_{\left( A_{2}\right) _{1}\left( A_{4}\right) _{1}}^{00}\right)
\end{eqnarray}%
\begin{eqnarray}
\left( I_{3}\right) _{A_{2}}^{1,3} &=&\frac{1}{2}\left( D_{\left(
A_{2}\right) _{1}}^{000}+D_{\left( A_{2}\right) _{1}}^{001}\right) \left(
D^{0000}+D^{0001}-D^{0010}-D^{0011}\right)  \notag \\
&&-\left[ D_{\left( A_{2}\right) _{1}\left( A_{4}\right) _{1}}^{00}\left(
D_{\left( A_{4}\right) _{0}}^{000}-D_{\left( A_{4}\right) _{0}}^{001}\right)
+D_{\left( A_{2}\right) _{1}\left( A_{4}\right) _{0}}^{00}\left( D_{\left(
A_{4}\right) _{1}}^{000}-D_{\left( A_{4}\right) _{1}}^{001}\right) \right] .
\end{eqnarray}%
\begin{equation}
\left( I_{3}\right) _{A_{2}}^{0,4}=\left( D_{\left( A_{2}\right)
_{1}}^{000}+D_{\left( A_{2}\right) _{1}}^{001}\right) ^{2}-4D_{\left(
A_{2}\right) _{1}\left( A_{4}\right) _{0}}^{00}D_{\left( A_{2}\right)
_{1}\left( A_{4}\right) _{1}}^{00}.
\end{equation}

Four-qubit invariant that quantifies the sum of three-way and four-way
correlations of triple $A_{1}A_{3}A_{4},$ reads as%
\begin{equation}
16N_{4,8}^{A_{1}A_{3}A_{4}}=16\left( 6\left\vert \left( I_{3}\right)
_{A_{2}}^{2,2}\right\vert ^{2}+4\left\vert \left( I_{3}\right)
_{A_{2}}^{3,1}\right\vert ^{2}+4\left\vert \left( I_{3}\right)
_{A_{2}}^{1,3}\right\vert ^{2}+\left\vert \left( I_{3}\right)
_{A_{2}}^{4,0}\right\vert ^{2}+\left\vert \left( I_{3}\right)
_{A_{2}}^{0,4}\right\vert ^{2}\right) ,
\end{equation}%
while degree-eight invariant that detects genuine four-body entanglement of
a four-qubit state is given by%
\begin{equation}
I_{4,8}=3\left( \left( I_{3}\right) _{A_{2}}^{2,2}\right) ^{2}-4\left(
I_{3}\right) _{A_{2}}^{3,1}\left( I_{3}\right) _{A_{2}}^{1,3}+\left(
I_{3}\right) _{A_{2}}^{4,0}\left( I_{3}\right) _{A_{2}}^{0,4}.  \label{i4inv}
\end{equation}%
Invariant $I_{4,8}$ is expressed here as a function of $A_{1}A_{3}A_{4}$
invariants. Being independent of the choice of focus qubit, $I_{4,8}$ can
also be written as a function of $A_{1}A_{2}A_{3}$ invariants or $%
A_{1}A_{2}A_{4}$ invariants. Sets $\left\{ \left( I_{3,4}\right)
_{A_{4}}^{4-m,m}:m=0\text{ to }4\right\} $ for qubits $A_{1}A_{2}A_{3}$ and $%
\left\{ \left( I_{3,4}\right) _{A_{3}}^{4-m,m}:m=0\text{ to }4\right\} $ for
the triple $A_{1}A_{2}A_{4}$, can be constructed from two-qubit invariants
of properly selected pair of qubits \cite{shar16}. Four-tangle based on
degree-eight invariant is defined \cite{shar16} as%
\begin{equation}
\tau _{1|2|3|4}=16\left\vert 12\left( I_{4,8}\right) \right\vert .
\label{genuine}
\end{equation}

\end{document}